\documentclass[9pt,twocolumn,twoside]{osajnl}

\journal{ol} 
\usepackage{graphicx}
\usepackage{dcolumn}
\usepackage{lipsum}
\usepackage{bm}
\usepackage{tabu}
\usepackage[utf8]{inputenc}
\usepackage[T1]{fontenc}

\usepackage{color}

\usepackage{comment}
\usepackage{soul}
\usepackage{siunitx}

\setboolean{shortarticle}{true}

\title{Experimental generation of cylindrical vector modes via an astigmatic mode converter}

\author[1]{Tatiana Román-Valenzuela}
\author[2]{Valeria Rodríguez-Fajardo}
\author[3]{Xiao Bo-hu}
\author[1*]{Carmelo Rosales-Guzm\'an}
\affil[1]{Centro de Investigaciones en Óptica, A.C., Loma del Bosque 115, Colonia Lomas del campestre, 37150 León, Gto., Mexico.}
\affil[2]{Department of Physics and Astronomy, Colgate University, Hamilton, NY 13346, United States of America.}
\affil[3]{Key Laboratory of Optical Field Manipulation of Zhejiang Province,Department of Physics, Zhejiang Sci-Tech University, Hangzhou, 310018, China}
\affil[*]{Corresponding author: carmelorosalesg@cio.mx}

\begin{abstract}
In this work, we propose and demonstrate experimentally a compact technique for the generation of cylindrical vector beams, based on a Michelson interferometer and a $\pi$-astigmatic mode converter, capable of inverting the topological charge of higher-order Laguerre-Gauss beams. Compared to previously demonstrated methods, this is relatively easy to align, and very compact. In addition, it generalises the concept of astigmatic mode conversion, commonly associated with scalar beams, to vector beams with non-homogeneous polarisation distribution. We anticipate that many of the applications based on Michelson interferometers will benefit from the unique properties of vector beams.  
\end{abstract}

\setboolean{displaycopyright}{true}

\begin{document}
\maketitle

Vector vortex beams, commonly regarded as a nonseparable superposition of the spatial and polarisation degrees of freedom, have become a topic of late, pioneering a wide diversity of applications, which have impacted research fields such as optical tweezers, optical metrology, optical communications, amongst many others  \cite{Zhan2009,Rosales2018Review,Yuanjietweezers2021,Ndagano2018}. They have been also identified as the classical analogous of quantum entangled states, an analogy that has coined them the controversial term classically-entangled beams, and has allowed the adaptation of many of the tools initially developed to describe the quantum word, to the classical word \cite{toninelli2019,forbes2019classically}. In addition, it has triggered the development of quantum-inspired technologies with vector beams \cite{Shen2022}. As such, in the last two decades, a variety of generation techniques have been proposed. Noteworthy, the advent of computer-controlled devices ignited a new era, the era of computer-generated holography, which has enabled the generation of arbitrary vector beams. Liquid crystal Spatial Light Modulators (SLMs) \cite{Davis2000,Moreno2012,Mitchell2017,Rong2014} and more recently Digital Micromirror Devices (DMDs) \cite{Ren2015,Mitchell2016,Scholes2019,Gong2014,hu2021generation} represent two of the most popular technologies. One of the main differences between SLMs and DMDs is the polarisation-dependent attribute of the first, which only allows the modulation of linear polarisation (typically horizontal) \cite{SPIEbook}, while the latter can modulate any polarisation state \cite{hu2021generation}. Hence, the generation of vector beams with an SLM requires independent manipulation of the transverse profile of both polarisation components, commonly employing interferometric arrays \cite{Maurer2007,Rosales2017,Liu2018,Mendoza-Hernandez2019,Galvez2012,Niziev2006}.  The use of DMDs simplifies these tasks, with the additional advantage of allowing higher refreshing rates \cite{Rosales2020,Perez-Garcia2022}. Given that vector beams are generated as the nonseparable superposition of the spatial and polarisation degrees of freedom, most interferometric techniques are restricted to the use of Mach-Zehnder or Sagnac interferometric arrays \cite{Volostnikov2013,Wang2007,Liu2012,XuXiang2018,Perez-Garcia2017}. 

\begin{figure*}[tb]
    \centering
    \includegraphics[width=.97\textwidth]{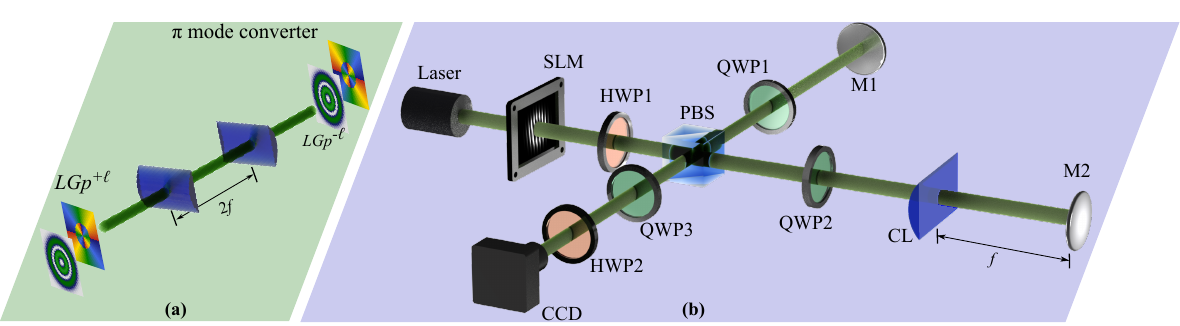}
    \caption{(a) A $\pi$-mode converter inverts the topological charge of an LG mode, from $\ell$ to $-\ell$. (b) Experimental setup for the generation of vector beams using a $\pi$-mode converter via a Michelson interferometer. A commercial diode-laser with horizontal polarisation, expanded and collimated, impinges on a Spatial Light Modulator (SLM), where an image containing the computer-generated hologram of the desired beam is displayed. A Half-Wave plate (HWP1) rotates the polarisation of the generated beam from horizontal to diagonal. The beam is then separated with a Polarising Beam Splitter (PBS) into its horizontal and vertical polarisation components, which travel along orthogonal paths. Two mirrors return the beam towards the PBS forming a Michelson interferometer. A Quarter Wave-Plate (QWP) is placed on the path of each beam, to ensure they both exit to the face adjacent to the input port, where the vector beam is generated. A Cylindrical Lens (CL) of focal length $f$ placed in any of the two paths and at a distance f from the mirror, inverts the topological charge of the input mode from $\ell$ to $-\ell$ after going through the same cylindrical lens twice. A QWP located at the output port of the interferometer transforms the vector beam from the linear to the circular bases. A second HWP (HWP2) enables the rotation of the polarisation distribution of the vector beam, and finally a Charge-Coupled Device (CCD) camera is used to measure the necessary intensities.}
    \label{fig:setup}
\end{figure*}

Here we demonstrate a method for generating vector beams employing a Michelson interferometer, one of the most famous that has given rise to various applications, such as Optical Coherence Tomography (OCT) \cite{OCT}. Nonetheless, given its configuration, and the fact that both beams inside the interferometer travel twice along the same optical path, it does not allow in principle to generate vector beams. Therefore, it has never been used for this purpose, even though many of the applications based on such interferometer would benefit from the additional properties the polarisation degree of freedom offers. Our technique requires the use of an astigmatic mode converter using a pair of cylindrical lenses \cite{Allen1992,Padgett2002,Beijersbergen1993}. There are two types of such astigmatic mode converters, with a theoretical formulation detailed in \cite{Beijersbergen1993}. The first is the $\pi/2$-mode converter, which transforms Hermite-Gauss (HG) beams into Laguerre-Gauss (LG), as illustrated in Fig.~\ref{fig:setup} (a). This mode converter is implemented by placing two cylindrical lenses of identical focal length $f$ facing each other and separated by a distance $d$ given in terms of the focal length as $d=\sqrt{2}f $. The second mode converter, capable of inverting the topological charge of LG beams, can be implemented similarly but with a separation distance $d=2f$, as schematically shown in Fig.~\ref{fig:setup} (b).

To begin with, let's recall that cylindrical vector beams can be described mathematically as the weighted non-separable superposition of the spatial and polarisation degrees of freedom as \cite{Rosales2018Review},
\begin{equation}
U(\mathbf{r}) = \cos\theta\, u_R(\mathbf{r}) \mathbf{\hat{e}_R} + \sin\theta \,u_L(\mathbf{r}) e^{i\delta}\mathbf{\hat{e}_L},
\label{VB}
\end{equation}
where the functions $u_R(\mathbf{r})$ and $u_L(\mathbf{r})$ are in general two orthogonal spatial modes. Further, the unitary vectors $\mathbf{\hat{e}_R}$ and $\mathbf{\hat{e}_L}$ represent the right and left-handed circular polarisation. The parameter $\theta \in [0, \pi/2]$ is a weighing coefficient that allows a transition from scalar to vectorial. The parameter $\delta \in [0,\pi]$ is an intermodal phase that allows a transition from different polarisation states. Here we will restrict ourselves to the set of Laguerre-Gauss Vector (LGV) beams, mathematically expressed as,
\begin{equation}
U(\mathbf{r}) = \cos\theta LG_p^{\ell_1}(\mathbf{r}) \mathbf{\hat{e}_R} + \sin\theta LG_p^{\ell_2}(\mathbf{r}) e^{i\delta}\mathbf{\hat{e}_L},
\label{LGVB}
\end{equation}
where $LG_p^\ell$ represents an LG beam of radial index $p \in \mathbb{N}$ and azimuthal index $\ell \in \mathbb{Z}$. Even though in general $\ell_1\neq \ell_2$, here we will only consider the subset of LGV beams for which $\ell_2 = -\ell_1$, which provide a wide variety of vector beams that are routinely used in various applications. 

A schematic representation of the setup implemented to generate vector modes via a Michelson interferometer and a $\pi$ mode converter is illustrated in Fig.~\ref{fig:setup} (c). Here, an expanded and collimated beam with horizontal polarisation is passed through a transmissive Spatial Light Modulator (LC2012 SLM from Holoeye), with a spatial resolution of $1024\times768$ and a pixel size of 36 $\mu$m. The Computer-Generated Hologram (CGH) that corresponds to the desired LG beam, computed through Complex Amplitude Modulation (CAM) is displayed onto the SLM. The first diffraction order is then spatially filtered with the help of a telescope and a variable aperture placed at the middle of both lenses (none of these are shown in the scheme). The polarisation direction of the emerging beam is rotated to diagonal and sent to a modified Michelson interferometer that incorporates a polarising beam splitter. The beam splits into its horizontal and vertical polarisation components that propagate along the two arms of the interferometer. A cylindrical lens with focal length $f$ is inserted in one of the arms and placed at a distance $d=f$ from the mirror that reflects back the beam to the same cylindrical lens. As a result, the beam effectively travels through a $\pi$-mode converter, which inverts its topological charge, as illustrated in Fig.~\ref{fig:setup}(c). To optimise the input power, a Quarter wave-plate inserted in each arm transforms the horizontal polarisation to vertical and the vertical to horizontal to redirect both beams to the face the PBS adjacent to the input port. Careful alignment with the mirrors allows the overlap of both beams, thereby generating the vector beam. In addition, a QWP allows the transformation of the generated vector beam from the linear to the circular polarisation bases. The intermodal phase of the generated beams, the parameter $\delta$ in Eq. \ref{LGVB}, can be introduced with an HWP. The generated vector beams are characterised through Stokes polarimetry and a Charge-Coupled Device (CCD) camera as explained next. As a final comment, even though here we used an SLM for the generation of the input beam, this can be generated with any other available method, using for example a Digital Micromirror Device or a low-cost printed hologram, as in \cite{Torres-Leal2023}. 

\begin{figure}[tb]
    \centering
    \includegraphics[width=.45\textwidth]{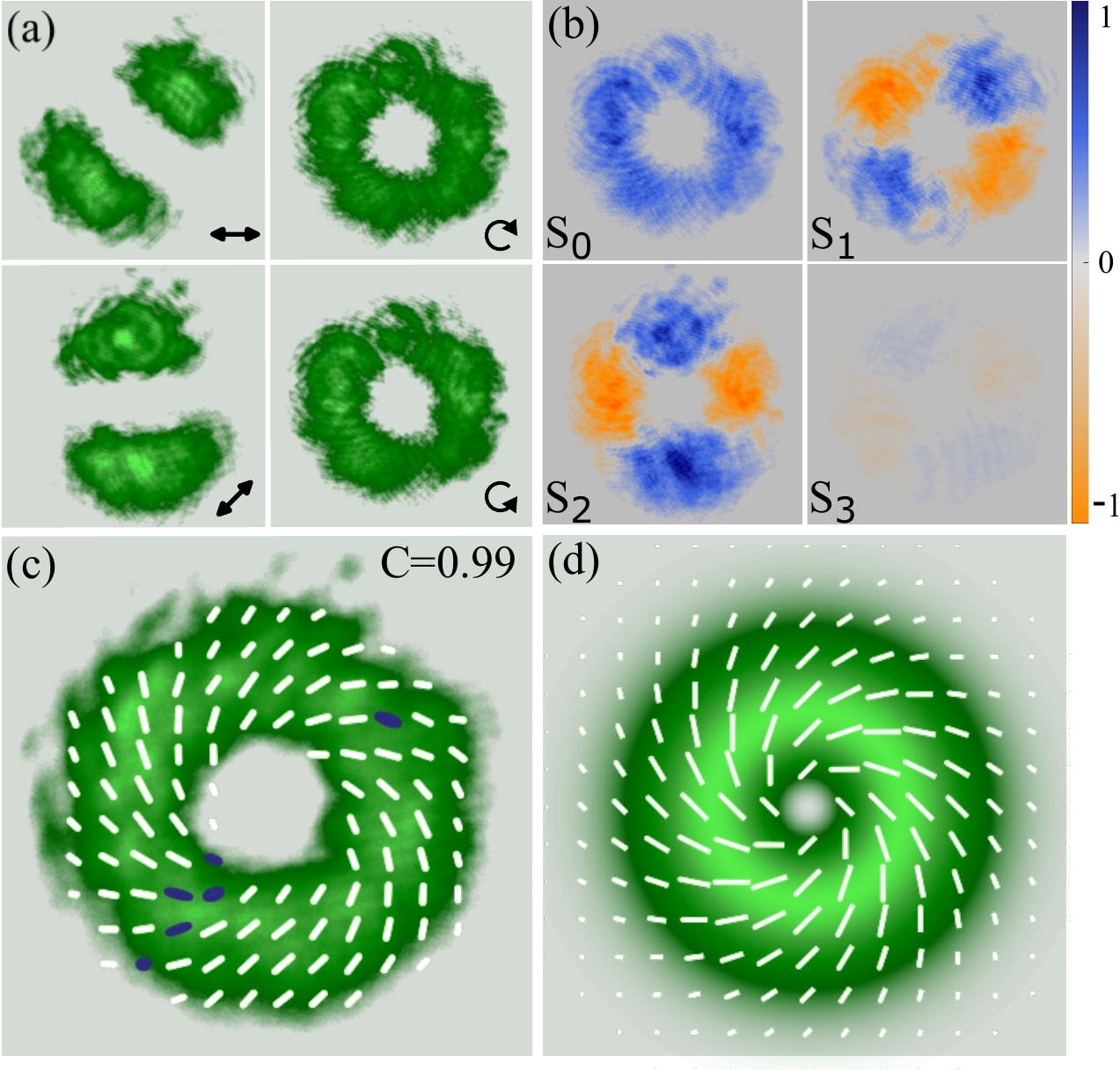}
    \caption{Example procedure to reconstruct polarisation states through Stokes polarimetry. (a) Experimentally measured intensities, with the bottom-left arrows indicating the direction of measured polarisation projection. (b) Stokes parameters reconstructed from the measured intensities. (c) Experimental intensity profile overlapped with the reconstructed polarisation distribution. (d) Theoretical intensity and polarisation distribution.}
    \label{fig:stokes}
\end{figure}

To corroborate the experimental accuracy of the beams generated with this technique, we reconstructed the transverse polarisation distribution of several CVB through Stokes polarimetry. This is based on a series of four intensity measurements \cite{rosales2021Mathieu}, such that the first two ($I_H$ and $I_D$) correspond to the projection of the vector beam onto a linear polariser oriented horizontally and diagonally, respectively, and the last two ($I_R$ and $I_L$) correspond to the right- and left-handed polarisation components of the vector beam, commonly measured through the combination of a QWP and a linear polariser. With these, the four Stokes parameters are calculated according to the relations,
\begin{equation}
    \label{eq:stokesParams}
    \begin{split}
        &S_0 = I_R + I_L , \hspace{10mm} S_1 = I_H - S_0, \\
        &S_2 = I_D - S_0, \hspace{10mm} S_3 = I_R - I_L,\\
    \end{split}
\end{equation}
from which the transverse polarisation distribution can be reconstructed \cite{Goldstein2011}. By way of example, Fig.~\ref{fig:stokes}(a) shows the measured intensities for a Laguerre-Gauss vector beam with spiral polarisation, for which, $\ell=1$, $\theta=\pi/4$ and $\delta=\pi/2$. The arrows in the bottom right indicate the corresponding measured polarisation projection. The Stokes parameters reconstructed from these intensities are shown in Fig.~\ref{fig:stokes} (b). Notice that the Stokes parameter $S_3$ has a near null value over the profile, indicating that $I_R$ and $I_L$ have the same intensity value over the whole plane. The transverse intensity profile overlapped with the polarisation distribution (represented by white lines) reconstructed from the Stokes parameters is illustrated in Fig.~\ref{fig:stokes}(c). By comparison, Fig.~\ref{fig:stokes}(d) shows its theoretical counterpart, computed from Eq. \ref{LGVB}. Notice the similitude between the measured and theoretical polarisation distributions.

As an additional measurement to corroborate the quality of the vector beams generated through this technique, we computed their Vector Quality Factor (VQF), also known as concurrence $C$, a quantity inspired in quantum mechanics tools, that has gained popularity in recent time \cite{Ndagano2016,Zhao2020,Selyem2019}. The VQF assigns values in the interval [0,1], 0 for scalar beams, and 1 for pure vector beams. In the latter, the spatial and polarisation degrees of freedom are maximally coupled. Here, for simplicity, we implemented a recently proposed version which is basis-independent and can be directly computed from the Stokes parameters \cite{Selyem2019,rosales2021Mathieu}. Mathematically, $C$ can be computed through the expression,
\begin{equation}
C=\sqrt{1-\left(\frac{\mathbb{S}_1}{\mathbb{S}_0} \right)^2-\left(\frac{\mathbb{S}_2}{\mathbb{S}_0} \right)^2-\left(\frac{\mathbb{S}_3}{\mathbb{S}_0} \right)^2},
\label{concurrence}
\end{equation}
where the parameters $\mathbb{S}_i$ represent the integral over the entire transverse plane of the Stokes parameter, that is, 
\begin{equation}
\mathbb{S}_i=\iint_{-\infty}^\infty S_{i}\, dA, \qquad i=0,1,2,3.
\end{equation}

\begin{figure}[tb]
    \centering
    \includegraphics[width=.48\textwidth]{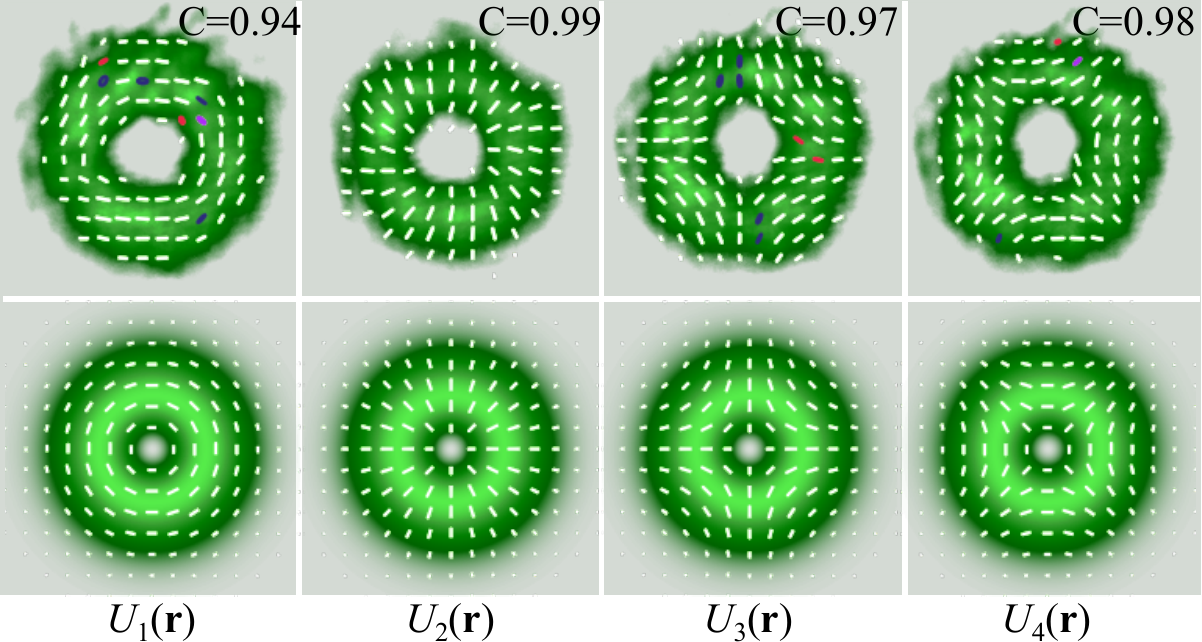}
    \caption{Intensity overlapped with the transverse polarisation distribution of a set of LG vector beams given by the different combinations of the topological charge $\ell$. The top panels correspond to the measured cases, whereas the bottom to the theory. The number indicates the VQF of each beam.}
    \label{fig:Bell}
\end{figure}

As a first example, to illustrate the capabilities of our technique, we generated the set of vector four beams, given by $p=0$ and all possible combinations of $\ell=\pm1$. These are commonly known as Bell states\cite{Shen2022}, are of great relevance in optical communications as they are eigenmodes of both free-space and optical fibres, and are described mathematically by the expressions,
\begin{equation}\centering
\begin{split}
&U_1(\mathbf{r}) = \frac{1}{\sqrt{2}}\left[ LG_0^{1}(\mathbf{r}) \mathbf{\hat{e}_R} +  LG_0^{-1}(\mathbf{r})\mathbf{\hat{e}_L}\right],\\
&U_2(\mathbf{r}) = \frac{1}{\sqrt{2}}\left[ LG_0^{1}(\mathbf{r}) \mathbf{\hat{e}_R} -  LG_0^{-1}(\mathbf{r})\mathbf{\hat{e}_L}\right],\\
&U_3(\mathbf{r}) = \frac{1}{\sqrt{2}}\left[ LG_0^{-1}(\mathbf{r}) \mathbf{\hat{e}_R} +  LG_0^{1}(\mathbf{r})\mathbf{\hat{e}_L}\right],\\
&U_4(\mathbf{r}) = \frac{1}{\sqrt{2}}\left[ LG_0^{-1}(\mathbf{r}) \mathbf{\hat{e}_R} - LG_0^{1}(\mathbf{r})\mathbf{\hat{e}_L}\right],
\end{split}
\label{eq:bell}
\end{equation}
The measured intensity and polarisation distribution are shown in Fig.~\ref{fig:Bell}, experiment on top and the theory on the bottom. For the sake of clarity, each beam is labelled according to Eq.\ref{eq:bell}. In each case, the measured VQF is shown in the top-right corner, showing very high-quality values, close to the ideal value $C=1$.

\begin{figure}[tb]
    \centering
    \includegraphics[width=.48\textwidth]{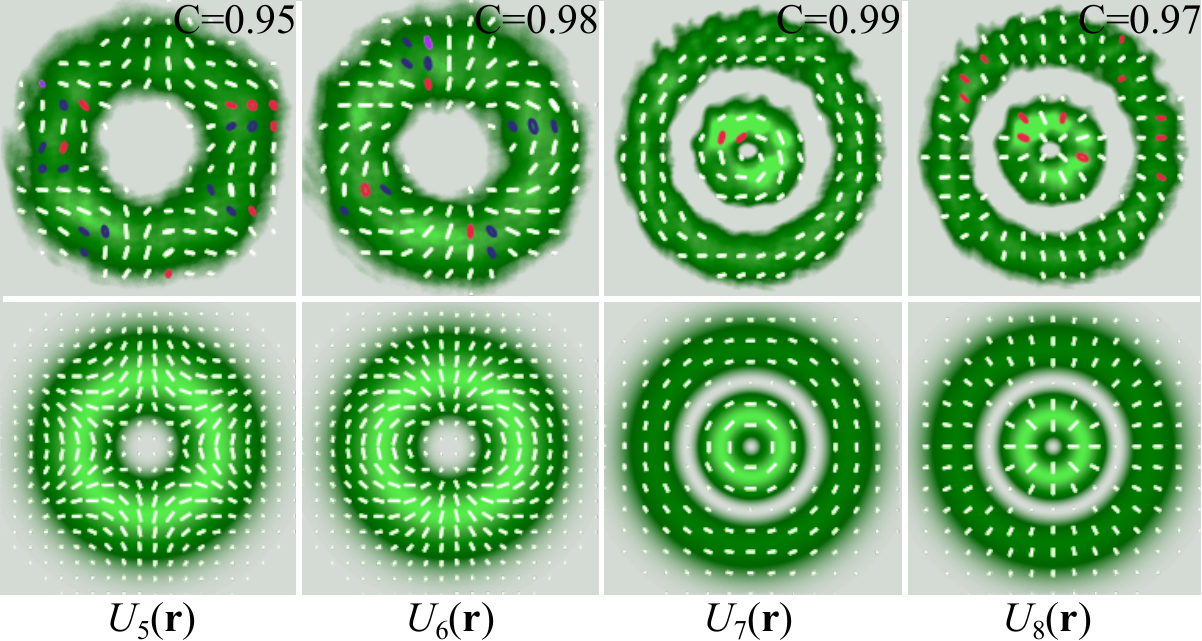}
    \caption{Intensity distribution overlapped with the transverse polarisation distribution of LGV modes formed by the non-separabale superposition of LG beams with parameters (a) and $p=0$ and $\ell=\pm 2$.}
    \label{fig:LG}
\end{figure}

We further demonstrated the potential of our technique by generating higher-order LG beams with $|\ell|>1$ and $p>0$. A set of representative examples given by the expressions,
\begin{equation}\centering
\begin{split}
&U_5(\mathbf{r}) = \frac{1}{\sqrt{2}}\left[ LG_0^{2}(\mathbf{r}) \mathbf{\hat{e}_R} +  LG_0^{-2}(\mathbf{r})\mathbf{\hat{e}_L}\right],\\
&U_6(\mathbf{r}) = \frac{1}{\sqrt{2}}\left[ LG_0^{-2}(\mathbf{r}) \mathbf{\hat{e}_R} +  LG_1^{2}(\mathbf{r})\mathbf{\hat{e}_L}\right],\\
&U_7(\mathbf{r}) = \frac{1}{\sqrt{2}}\left[ LG_1^{1}(\mathbf{r}) \mathbf{\hat{e}_R} +  LG_1^{1}(\mathbf{r})\mathbf{\hat{e}_L}\right],\\
&U_8(\mathbf{r}) = \frac{1}{\sqrt{2}}\left[ LG_1^{1}(\mathbf{r}) \mathbf{\hat{e}_R} - LG_1^{-1}(\mathbf{r})\mathbf{\hat{e}_L}\right],
\end{split}
\label{eq:higher}
\end{equation}
are shown in Fig.~\ref{fig:LG}, experiment on top and theory on the bottom. Here, similarly to the previous case, each panel is labelled with its corresponding vector beam from Eq. \ref{eq:higher}. Notice the high similarity between the measured polarisation distribution and their theoretical counterpart. In a similar way to the previous example, we also determined their VQF, which is also shown in the top-right insets. The measured values are all close to one, evincing their high quality.

In conclusion, in this manuscript we demonstrated a novel technique for the generation of cylindrical vector beams, which is based on a modified Michelson interferometer that incorporates a polarising beam splitter (PBS). The technique relies on the well-known $\pi$-astigmatic mode converter, formed by two cylindrical lenses separated by a distance $2f$, which inverts the topological charge of Laguerre-Gauss (LG) beams. Here, an input scalar LG beam with diagonal polarisation is inserted into the Michelson interferometer and separated into its two polarisation components. The $\pi$-mode converter, implemented by passing one of the LG beams twice over the same cylindrical lens, reverses its topological charge. Hence, upon recombination of both scalar beams with orthogonal polarisation at the PBS, a vector beam is generated. Even though the Michelson interferometer is one of the most popular, it has been mainly associated to interferometric experiments with scalar beams. Here we put forward and advanced version of a Michelson interferometer, which in combination with a $\pi$ mode converter, allows the generation of so-called vector beams, states of light with non-homogeneous polarisation distribution. We anticipate that our proposed device will be of great relevance in advancing existing applications, such as optical coherence tomography, but also in the development of new. For example, in optical metrology, it could be use as an alternative tool to measure via polarisation, the index of refraction or to perform surface profilometry. This device will also allow the generation of partially coherent vector beams, a topic that has gained popularity in recent time.


\section*{Funding}
This research was supported by Zhejiang Provincial Natural Science Foundation of China under Grant No. LQ23A040012.
\section*{Disclosures}
The authors declare that there are no conflicts of interest related to this article.


\end{document}